\documentclass[12pt]{article}

\usepackage{amsmath,amssymb,color,graphics}
\usepackage[
colorlinks=true,backref,pagebackref,pdfa]{hyperref} 
\def\a{\alpha}
\def\b{\beta}
\def\d{\delta}
\def\g{\gamma}
\def\m{\mu}
\def\n{\nu}
\def\eps{\epsilon}
\def\ve{\varepsilon}
\def\vt{\vartheta}

\def\vt{\vartheta}

\usepackage{amsfonts}

\begin{document}

\title{On Freud's superpotential in General
  Relativity and in Einstein-Cartan theory}

\author{Christian G.\ B\"ohmer\footnote{c.boehmer@ucl.ac.uk}~$^1$ and
  Friedrich W.\ Hehl\footnote{hehl@thp.uni-koeln.de}~$^{2}$\\[2ex]
  $^{1}$Department of Mathematics, University College London\\
  Gower Street, London, WC1E 6BT, UK\\[1ex]
  $^{2}$Institute for Theoretical Physics, University of Cologne\\
  Z\"ulpicher Str.\ 77, 50937 K\"oln, Germany\\
  and Department of Physics \& Astronomy, University of Missouri\\
  Columbia, MO 65201, USA}

\date{}
\maketitle
\begin{abstract}
  The identification of a suitable gravitational energy in theories of
  gravity has a long history, and it is well known that a unique
  answer cannot be given. In the first part of this paper we present a
  streamlined version of the derivation of Freud's superpotential in
  general relativity. It is found if we once integrate the
  gravitational field equation by parts. This allows us to extend
  these results directly to the Einstein-Cartan theory. Interestingly,
  Freud's original expression, first stated in 1939, remains valid
  even when considering gravitational theories in Riemann-Cartan or,
  more generally, in metric-affine spacetimes.

\hfill {\begin{footnotesize}\it file
      BoehmerHehl\_Freud\_17.tex, 08 Dec 2017\end{footnotesize}}
\end{abstract}

%
\section{Introduction}\label{Intro}

\subsection{Gravity and energy}

General relativity does not lend itself naturally to the definition
of a gravitational energy. This led to the formulation of energy
via pseudo-tensors or an energy-momentum density complex, see Chen,
Nester, and Tung \cite{Chen:2015vya} for a brief historical account;
earlier contributions, chronologically ordered, include
\cite{ThirringWallner,Nester:1981,Dubois,Frauendiener,Szabados,Itin:2001bp,Itin:2001xz,Nester:2004,So:2008kr}.
These objects appear more or less naturally when studying
gravitational field equations, depending on one's approach. In the
present paper we proceed by revisiting the Einstein field equation
written in exterior calculus. This has the great advantage that a
single integration by parts allows us to define a conserved quantity,
which can be related directly with Freud's superpotential
\cite{Freud}. Within the context of general relativity, most of these
results are known and can be found in the literature.

Our approach allows us to extend these results smoothly to other
gravitational theories. First, it will be applied to Einstein-Cartan
theory where we will derive the analogue superpotential in the
presence of torsion. Second, we show how the formalism naturally
includes all metric-affine gravity theories. It is particularly
noteworthy that Freud's original ``affine tensor density'' applies to
all such theories despite Freud's proof being given only in the case
of general relativity.

\subsection{Notation and conventions}

For exterior calculus and its application to field theory, we
recommend, as examples, Bamberg \& Sternberg \cite{Bamberg:1990},
Agricola \& Friedrich \cite{Agricola:2002}, and Scheck
\cite{Scheck:2012}. We take our corresponding conventions from
\cite{PRs} (see also \cite{Erice95}). For tensor calculus, see
Schouten \cite{Schouten:1954,Schouten:1989} and, for the generalized
Kronecker symbols and the $\epsilon$-system, in particular Sokolnikoff
\cite{Sokolnikoff}.

The coframe one-form is denoted by $\vt^\a$, with Greek (anholonomic)
indices $\a,\b,\g,...=0,1,2,3$. The vector frame $e_\b$ is dual to the
coframe, that is, the interior product of the frame as applied to the
coframe reads $e_b\rfloor \vt^\a=\d^\a_\b$. (Holonomic) coordinate
indices are written in Latin, namely $i,j,k,...=0,1,2,3$. A $p$-form
$\Psi$ expanded with respect to the anholonomic basis is given by
\begin{equation}
  \Psi = \frac{1}{p!} \Psi_{\alpha_1 \ldots \alpha_p} \vt^{\alpha_1}
  \wedge   \cdots \wedge \vt^{\alpha_p} \,.\label{pform}
\end{equation}

The $\eta$-basis is defined in the conventional way, see \cite{PRs}:
If we take the interior product $\rfloor$ of an arbitrary frame
$e_\alpha$ with the {\em metric volume element} 4-form
${\eta}:=\sqrt{|\det g_{\mu\nu}|}\,\vt^0\wedge\vt^1\wedge\vt^2\wedge\vt^3$,
then we find a 3-form ${\eta}_\alpha$; if we contract again, we find a
2-form ${\eta}_{\alpha\beta}$, etc.:
\begin{eqnarray}
  {\eta}_{\alpha} &:=& e_{\alpha} \rfloor
  {\eta} = \frac{1}{3!}\,
  {\eta}_{\alpha\beta\gamma\delta}\, \vartheta^{\beta} \wedge
  \vartheta^\gamma \wedge \vartheta^{\delta}={}^\star\vt_\a\,, \\ 
  {\eta} _{\alpha\beta}& := & e_{\beta} \rfloor
  {\eta} _{\alpha} =
  \frac{1}{2!}\,{\eta}_{\alpha\beta\gamma\delta} \, \vartheta
  ^{\gamma}\wedge \vartheta ^{\delta} ={}^\star(\vt_\a\wedge\vt_\b)\,, \\ 
  {\eta}_{\alpha\beta\gamma}& := & e_{\gamma} \rfloor
  {\eta}_{\alpha\beta}=
  \frac{1}{1!}\,{\eta}_{\alpha\beta\gamma\delta}\, \vartheta^\delta
 ={}^\star(\vt_\a\wedge\vt_\b\wedge\vt_\g)\,,
 \label{eta3i} \\
       {\eta}_{\alpha\beta\gamma\delta}\, & \; := & 
  e_\delta\rfloor{\eta}_{\alpha\beta\gamma}=e_{\delta}\rfloor
  e_\gamma\rfloor e_\beta\rfloor
  e_\alpha\rfloor{\eta} ={}^\star(\vt_\a\wedge\vt_\b\wedge\vt_\g\wedge\vt_\d)\,.
\end{eqnarray}
Similar formulas are valid for the $\epsilon$-system, where
$\epsilon_{\a\b\g\d}=\pm 1,0$ is the Levi-Citita symbol, see
\cite{PRs,Birkbook}. The Hodge star $^\star$ is then substituted by
the diamond star $^\diamond$. This system will be used in
metric-affine spacetimes.

In Riemannian geometry a tilde above an
expression $\tilde{M}$ is used. This means such quantities are based
on the Christoffel symbol components without torsion or
nonmetricity. In Riemann-Cartan space the corresponding quantities
are denoted {\it without} a tilde. The torsion 2-form and the
curvature 2-forms read, respectively,
\begin{equation}\label{torcurv}
  T^\a:=D\vt^\a=d\vt^a +\Gamma_\b{}^\a\wedge\vt^\b\,,\quad
  {R}_\alpha{}^\beta=d{\Gamma}_\alpha{}^\beta-
  {\Gamma}_\alpha{}^\gamma
  \wedge{\Gamma}_\gamma{}^\beta\,.
\end{equation}

Einstein's gravitational constant and the speed of light are put to
one: $\kappa=1,\,c=1$.

\section{Freud's superpotential in general relativity}\label{GR}

\subsection{General relativity in exterior calculus}

It is possible to express Einstein's field equation \cite{Meaning} in
the framework of the calculus of exterior differential forms, see
Misner, Thorne, and Wheeler \cite{MTW}, Trautman \cite{Trautman},
Thirring \& Wallner \cite{ThirringWallner,Thirring}, and Kopczy\'nski
\cite{Kopczynski:1990af}. Let $\tilde{R}_\alpha{}^\beta$ be the
curvature 2-form of the (pseudo-)Riemannian spacetime,
\begin{equation}\label{curv}
  \tilde{R}_\alpha{}^\beta=d\tilde{\Gamma}_\alpha{}^\beta-
\tilde{\Gamma}_\alpha{}^\gamma
  \wedge\tilde{\Gamma}_\gamma{}^\beta\,,
\end{equation}
with $\tilde{\Gamma}_\alpha{}^\beta$ as the Levi-Civita connection
1-form. We call the Einstein 3-form $\tilde{G}_\alpha$. Then the
Einstein equation reads \cite{Trautman}:
\begin{equation}\label{Einstein}
  \tilde{G}_\alpha := \frac{1}{2}\,{\eta}_{\alpha\beta\gamma}\wedge
  \tilde{R}^{\beta\gamma}=  {\cal T}_\alpha\,.
\end{equation}
Here the source term ${\cal T}_\alpha$ represents the symmetric
Hilbert energy-momentum 3-form of matter. 

To recover the standard form of the Einstein field equations, where the
Einstein tensor has the form of a trace-reversed Ricci tensor, one can
isolate the components of the Einstein 3-form using (\ref{pform}) and
(\ref{eta3i}). Finally, applying the Hodge dual to obtain the
1-form whose components will be the usual Einstein tensor components.

If one desires to introduce a superpotential \`a la Freud
\cite{Freud}, then one has to isolate in (\ref{Einstein}) a leading
term of the general form $\sim d({\eta}_{..}\wedge \tilde{\Gamma})$. In
exterior calculus, it is straightforward to see how to achieve this. We
substitute (\ref{curv}) into (\ref{Einstein}):
\begin{equation}\label{subst}
  \frac{1}{2}\,{\eta}_{\alpha\beta\gamma}\wedge
  \left(d\tilde{\Gamma}^{\beta\gamma}-\tilde{\Gamma}^{\beta\delta} \wedge
    \tilde{\Gamma}_\delta{}^\gamma \right)=  {\cal T}_\alpha\,.
\end{equation}
The next step is now particularly simple, we integrate by parts and
find immediately
\begin{equation}\label{part}
  d\,\big\{\!-\frac{1}{2}\,{\eta}_{\alpha\beta\gamma}\wedge
    \tilde{\Gamma}^{\beta\gamma}\big\}+
  \frac{1}{2}\,\left(d{\eta}_{\alpha\beta\gamma}\right)\wedge
  \tilde{\Gamma}^{\beta\gamma}
  -\frac{1}{2}\,{\eta}_{\alpha\beta\gamma}\wedge\tilde{\Gamma}^{\beta\delta}
  \wedge \tilde{\Gamma}_\delta{}^\gamma =  {\cal T}_\alpha\,.
\end{equation}
It is possible to manipulate the second term on the left-hand side of
this equation: We recall that in a Riemannian space
$\tilde{D}{\eta}_{\alpha\beta\gamma}=0$. Thus,
\begin{align}
  0 &= d{\eta}_{\alpha\beta\gamma}
  -\tilde{\Gamma}_\alpha{}^\delta\wedge{\eta}_{\delta\beta\gamma}
  -\tilde{\Gamma}_\beta{}^\delta\wedge{\eta}_{\alpha\delta\gamma}
  -\tilde{\Gamma}_\gamma{}^\delta\wedge{\eta}_{\alpha\beta\delta}
  \nonumber \\ &=
  d{\eta}_{\alpha\beta\gamma}
  -3\,\tilde{\Gamma}_{[\alpha}{}^\delta\wedge{\eta}_{\beta\gamma]\delta}\,,
\end{align}
which implies
\begin{equation}\label{simplify}
d{\eta}_{\alpha\beta\gamma}=
3\,\tilde{\Gamma}_{[\alpha}{}^\delta\wedge{\eta}_{\beta\gamma]\delta}\,.
\end{equation}
We substitute (\ref{simplify}) into (\ref{part}) and find, after
some algebra,
\begin{equation}\label{part1}
  d\,\big\{\!-\frac{1}{2}\,{\eta}_{\alpha\beta\gamma}\wedge
    \tilde{\Gamma}^{\beta\gamma}\big\}-\eta_{\b\g[\a}\wedge
\tilde{\Gamma}_{\d]}{}^\b\!\wedge\!
\tilde{\Gamma}^{\d\g}
  =  {\cal T}_\alpha\,.
\end{equation}

One is now led to define the Freud 2-form
\cite{Freud}\footnote{According to W.\ Kopczy\'nski \cite{Wojtek},
  Trautman, around 1975, taught about gravitational radiation at
  Warsaw University.  There W.K.\ learned {}from him about the forms
  $\tilde{\cal F}_\alpha$ and $\tilde{t}_\alpha$.  Nurowski remembers that
  W.K.\ lectured about these forms in the early eighties. Later
  Trautman spoke about them in Erice and, probably, {}from Erice it
  went somehow to Vienna. These forms then appeared in papers of
  Thirring.  Incidentally, Trautman called $\tilde{F}_\alpha$ Freud's
  superpotential.}
\begin{equation}\label{Freud2-form}
  \tilde{\cal F}_\alpha:=-\frac{1}{2}\,{\eta}_{\alpha\beta\gamma}\wedge
  \tilde{\Gamma}^{\beta\gamma}
\end{equation} 
and the gravitational energy-momentum, the Trautman--Sparling 3-form
\cite{Sparling},
\begin{equation}\label{Sparling3-form}
 \tilde{t}_\alpha:=\eta_{\b\g[\a}\wedge\tilde{\Gamma}_{\d]}{}^\b\!\wedge\!
\tilde{\Gamma}^{\d\g}\,.
\end{equation}
Then we can rewrite (\ref{part1}) simply as
\begin{equation}\label{Einstein1}
  d\tilde{\cal F}_\alpha=\tilde{t}_\alpha + {\cal T}_\alpha \,.\end{equation}

Equation (\ref{Einstein1}) has been derived earlier also by
Frauendiener \cite{Frauendiener}, for example. However, he finds it
``somewhat startling to see the Einstein tensor to show up in this
connection'' \cite[page L239]{Frauendiener}. In our deduction, the
relation to the Einstein 3-form is apparent and quite natural. A similar
observation can be found in \cite{Chen:2015vya} where it was pointed
out that this derivation of the superpotential is in ``remarkable constrast''
to the tensor calculus approach. 

Our main result is respresented by the equation (\ref{Einstein1}),
together with the definitions (\ref{Freud2-form}) and (\ref{Sparling3-form}):
\begin{align}\label{sum}
  \begin{split}
    d\tilde{\cal F}_\alpha&=\tilde{t}_\alpha + {{\cal T}}_\alpha \\
    {\rm with}\quad
    \tilde{\cal F}_\alpha:=-\frac{1}{2}\,{\eta}_{\alpha\beta\gamma}\wedge
    \tilde{\Gamma}^{\beta\gamma}\,&\quad\text{and}\quad
    \tilde{t}_\alpha:=\eta_{\b\g[\a}\wedge\tilde{\Gamma}_{\d]}{}^\b\!\wedge\!
    \tilde{\Gamma}^{\d\g}\,.
  \end{split}
\end{align}
Clearly, because of Poincar\'e's lemma, $dd\tilde{\cal F}_\a=0$.
Accordingly, we find the conservation law
\begin{equation}\label{conservation}
  d\left(\tilde{t}_\alpha + {\cal T}_\alpha\right)=0 \,.
\end{equation}
This demonstrates that the non-tensorial 3-form $\tilde{t}_\alpha$
represents an energy complex of the gravitational field, see also
Schr\"odinger \cite{Schroedinger}.

It turns out that the Freud 2-form $\tilde{\cal F}_\alpha$ is not
exactly the object used by Freud in the derivation of the energy
complex but rather its Hodge dual. This is addressed in the following.

\subsection{Hodge dual of the Freud two-form}

In subsequent parts we will need the Hodge dual of the Freud two-form. By
applying the standard formulas for the Hodge star, we find
\begin{align}
  ^\star{\cal F}_\a &= {}^\star\! \left(-{\scriptstyle \frac{1}{2}} 
  {\eta}_{\alpha\beta\gamma}\wedge
  \tilde{\Gamma}^{\beta\gamma}  \right)=- \frac{1}{2}
  \,^\star\!\left(\eta_{\a\b\g\d}\,\vt^\d \wedge
  \tilde{\Gamma}^{\beta\gamma}\right)\nonumber
  \\ &= - \frac{1}{2} \eta_{\a\b\g\d}\,^\star\!\left(\vt^\d \wedge
  \tilde{\Gamma}^{\beta\gamma}\right) 
  = - \frac{1}{2} \eta_{\a\b\g\d}\,^\star\!\left(\vt^\d \wedge\vt^\ve\,
  \tilde{\Gamma}_\ve{}^{\beta\gamma}\right)\nonumber
  \\ &= - \frac{1}{2} \eta_{\a\b\g\d}\,\tilde{\Gamma}_\ve{}\,
^{\beta\gamma}\,^\star\!\left(\vt^\d\wedge\vt^\ve\right)\,.
  \label{HodgeF}
\end{align} 
Accordingly, we find the relatively compact formula
($\eta^{\d\zeta}=-\eta^{\zeta\d}$)
\begin{equation}\label{HodgeF1}
  ^\star{\cal F}_\a= \frac{1}{2} \eta_{\a\b\g\d}\,g^{\b\ve}
  \tilde{\Gamma}_{\zeta\ve}{}^{\gamma}\eta^{\zeta\d}\,.
\end{equation} 

The components of this expression are given implicitly by
${}^\star{\cal F}_\a = \frac{1}{2} {}^\star{\cal F}_{\m\n\a}$
$ \vt^\m \wedge\vt^\n$ or explicitly by
$ {}^\star{\cal F}_{\m\n\a} = e_\n\rfloor e_\m\rfloor ^\star{\cal
  F}_\a$.  Thus, Eq.~\eqref{HodgeF1} yields
\begin{eqnarray}
   {}^\star{\cal F}_{\m\n\a}& =&  e_\n\rfloor e_\m\rfloor(\frac{1}{2}
\eta_{\a\b\g\d}\,g^{\b\ve}
  \tilde{\Gamma}_{\zeta\ve}{}^{\gamma}\eta^{\zeta\d})=\frac{1}{2}\eta_{\a\b\g\d}
\,g^{\b\ve}
 \, \tilde{\Gamma}_{\zeta\ve}{}^{\gamma}\, e_\n\rfloor
 e_\m\rfloor\eta^{\zeta\d}
 \nonumber \\
&=& \frac{1}{2}\eta_{\a\b\g\d}\,g^{\b\ve}
 \, \tilde{\Gamma}_{\zeta\ve}{}^{\gamma}\, \eta^{\zeta\d}{}_{\mu\nu}
= \frac{1}{2}(\eta_{\a\b\g\d}\, \eta_{\mu\nu}{}^{\zeta\d})\,g^{\b\ve}
 \, \tilde{\Gamma}_{\zeta\ve}{}^{\gamma}\,.
\end{eqnarray}
It appears that the last expression becomes more transparent if we
raise the indices $\mu$ and $\nu$:
\begin{equation}\label{semi}
  {}^\star{\cal F}^{\m\n}{}_{\a}= \frac{1}{2}(\eta_{\a\b\g\d}\, 
  \eta^{\mu\nu\zeta\d})\,g^{\b\ve}
  \, \tilde{\Gamma}_{\zeta\ve}{}^{\gamma}= \frac{1}{2} 
  \d^{\,\mu\nu\zeta}_{\a\b\g\,}g^{\b\ve}
  \, \tilde{\Gamma}_{\zeta\ve}{}^{\gamma}\,.
\end{equation}
In the last transformation, we used the rules for the $\eta$-system,
see \cite[Eq.(40.5)]{Sokolnikoff} together with the generalized
Kronecker deltas.\footnote{The generalized Kroneckers and the related
  $\epsilon$-system are also discussed in \cite[Ch.A.1]{Birkbook}}
The object $ \d^{\,\mu\nu\zeta}_{\a\b\g\,}$ can be written as a determinant:
\begin{equation}\label{determinant}
 \d^{\,\mu\nu\zeta}_{\a\b\g\,}=\left|
\begin{array}{ccc}
\d^\m_\a   &\d^\n_\a   &\d^\zeta_\a  \\
\d^\m_\b   &\d^\n_\b   &\d^\zeta_\b  \\
\d^\m_\g   &\d^\n_\g   &\d^\zeta_\g  
 \end{array}\right|\,.
\end{equation}
If we substitute this into \eqref{semi}, we find eventually:
\begin{equation}
 {}^\star{\cal F}^{\m\n}{}_{\a}= \frac{1}{2}
\left|
\begin{array}{ccc}
\d^\m_\a   &\d^\n_\a   &\d^\zeta_\a  \\
g^{\m\ve}   &g^{\n\ve}   &g^{\zeta\ve}  \\
 \tilde{\Gamma}_{\zeta\ve}{}^{\m}  & 
\tilde{\Gamma}_{\zeta\ve}{}^{\n}   & 
\tilde{\Gamma}_{\zeta\ve}{}^{\zeta}  
 \end{array}\right|\,.
\end{equation}
This is the analogous result as the one found by Freud in holonomic
coordinates, as we will see.

Using the notation $\mathfrak{g}^{ik}:=\sqrt{|g|}g^{ik}$ Freud wrote
the expression of the superpotential in a compact form by using the
determinant
\begin{equation}
  \label{frakA}
  \mathfrak{A}^{in}{}_k
  =\frac{1}{2}\,\begin{array}{|ccc|}
  \d^i_k&\d^n_k&\d^m_k\\
  \mathfrak{g}^{ir}&\mathfrak{g}^{nr}&\mathfrak{g}^{m r}\\
  \tilde{\Gamma}_{m r}{}^i&\tilde{\Gamma}_{\mu r}{}^n&
  \tilde{\Gamma}_{m r}{}^\mu\end{array}=-\mathfrak{A}^{ni}{}_k\,.
\end{equation}
Clearly, ${}^\star{\cal F}^{\m\n}{}_{\a}$ and
$\mathfrak{A}^{in}{}_k/\sqrt{|g|}$ are the same mathematical object. The
former is expressed in arbitrary (anholonomic) frames, whereas the
latter is expressed in terms of curvilinear (holonomic) coordinates.

\section{Einstein-Cartan and metric-affine gravity}\label{ECandPG}

\subsection{Freud in Einstein-Cartan theory (EC)}\label{EC}
The field equations of EC , see, e.g., Trautman
\cite{Trautman,TrautmanEncyclopedia}, Blagojevi\'c et al.\
\cite{Reader}, or Obukhov \cite{Obukhov:2006,Obukhov:2018}, are
algebraic in $R^{\a\b}$ and $T^\a$, respectively. In spite of this, we
want to try to put them in a form that is reminiscent to the field
equations of Yang-Mills type:
\begin{alignat}{3}
  \underbrace{G_\a}_{\text{Einstein 3-form} } &:= 
  \frac{1}{2} \eta_{\a\b\g}\wedge R^{\b\g} = \mathfrak{T}_\a &&
  \quad \Longrightarrow \quad
  & d{\cal F}_\a -{t}_\a &=\mathfrak{T}_\a\,,\\
  \underbrace{P_{\a\b}}_{\text{Palatini 3-form}} &:= 
  \frac{1}{2} \eta_{\a\b\g}\wedge T^\g = \mathfrak{S}_{\a\b} &&
  \quad \Longrightarrow \quad
  & d(\cdots\!)-{s}_{\a\b}&=\mathfrak{S}_{\a\b}\,.
\end{alignat}
The sources on the right-hand sides of the two field equations are the
canonical 3-forms of energy-momentum and spin angular momentum of
matter, respectively.

Our deduction of a Freud 2-form will proceed in strict analogy to
the one in GR. We only have to drop the tildes. There is, however, one
difference: In a Riemann-Cartan space, the covariant exterior
derivatives of the $\eta$-forms do not vanish any longer. We rather
have, see \cite{PRs},
\begin{eqnarray}
  D\eta _{\alpha }& =&T^\d \wedge \eta _{\alpha \d }\,,\\
  D\eta _{\alpha \b }& =&
  T^\d \wedge \eta _{\alpha\b \d }\,,  \\
  D\eta _{\alpha \b\g }&= &T^\d \wedge \eta_{\a\b\g\d}\,, \label{D_eta3}\\
  D\eta _{\a\b\g\d }& =&0\,.
\end{eqnarray}
As a consequence of \eqref{D_eta3}, we now find, instead of the
Riemannian result \eqref{simplify}, the corresponding relation in a
Riemann-Cartan space as
\begin{equation}\label{simplify'}
  d {\eta}_{\alpha\beta\gamma}=
  3\,\Gamma_{[\alpha}{}^\delta\wedge {\eta}_{\beta\gamma]\delta} +
  T^\delta \wedge \eta_{\a\b\g\d}\,.
\end{equation}
This newly emerging torsion term is the basic difference between both
deductions. Interestingly, the torsion term will not enter the Freud
superpotential explicitly. 

Analogously to the above, we substitute the gravitational field strengths
curvature $R^{\b\g}$ and the torsion $T^\g$ into the two field
equations:
\begin{eqnarray}\label{subst1}
 \frac{1}{2}\, {\eta}_{\alpha\beta\gamma}\wedge
  \left(d{\Gamma}^{\beta\gamma}-{\Gamma}^{\beta\delta} \wedge
    \Gamma_\delta{}^\gamma \right)&=&   \mathfrak{T}_\alpha\,,\\
 \frac{1}{2}\,D\eta_{\a\b}= d({\scriptstyle \frac{1}{2}}
\eta_{\a\b})+\Gamma_{[\a}{}^\g\wedge\eta_{\b]\g}&=&\mathfrak{S}_{\a\b}\,.\label{subst2}
\end{eqnarray}
We partially integrate the first term of \eqref{subst1} and find immediately,
\begin{equation}\label{partx}
  d\,(\,\underbrace{-\frac{1}{2}\, {\eta}_{\alpha\beta\gamma}\wedge
    \Gamma^{\beta\gamma}}_{{\cal F}_\a}\,)+
  \frac{1}{2}\,\left(d {\eta}_{\alpha\beta\gamma}\right)\wedge
  \Gamma^{\beta\gamma}
  -\frac{1}{2}\, {\eta}_{\alpha\beta\gamma}\wedge\Gamma^{\beta\delta}
  \wedge \Gamma_\delta{}^\gamma =   \mathfrak{T}_\alpha\,.
\end{equation}
Now we substitute \eqref{simplify'} into \eqref{partx} and collect the
terms containing the $\Gamma$'s quadratically:
\begin{equation}\label{party}
  d\,(\,-\frac{1}{2}\, {\eta}_{\alpha\beta\gamma}\wedge
    \Gamma^{\beta\gamma})-\eta_{\b\g[\a}\wedge\Gamma_{\d]}{}^\b\wedge\Gamma^{\d\g}
+{\scriptstyle \frac 12}\eta_{\a\b\g\d}\,\Gamma^{\b\g}
\wedge T^\d =   \mathfrak{T}_\alpha\,.
\end{equation}
Following the GR calculation closely, we define the first term as the Freud 2-form 
\begin{equation}\label{Freud2-form'}
{{\cal F}_\alpha} :=-{\scriptstyle \frac{1}{2}} 
{\eta}_{\alpha\beta\gamma}\wedge\Gamma^{\beta\gamma}\,
\end{equation}
and, enriched by a torsion term, the gravitational energy-momentum
3-form
\begin{equation}\label{3-form'}
  t_\alpha:=\eta_{\b\g[\a}\wedge\Gamma_{\d]}{}^\b\wedge\Gamma^{\d\g}
-{\scriptstyle \frac 12}\eta_{\a\b\g\d}\,\Gamma^{\b\g}
\wedge T^\d\,.
\end{equation}
Thus, the first field equation of EC can simply be written as
\begin{equation}\label{First}
\boxed{  d\,{\cal F}_\alpha-t_\alpha = \mathfrak{T}_\alpha \,.}
\end{equation}

The second field equation is already in its final form. However, for
compactness we introduce the gravitational spin 3-form
\begin{equation}\label{gravSpin}
s_{\a\b}:=-\Gamma_{[\a}{}^\g\wedge\eta_{\b]\g}.
\end{equation}
Then the second field equation of EC reads as follows:
\begin{equation}\label{Second}
\boxed{  d({\scriptstyle \frac{1}{2}}
  \eta_{\a\b})-s_{\a\b}=\mathfrak{S}_{\a\b}\,.}
\end{equation}

If we define the energy-momentum and spin complexes
\begin{equation}
\check{\mathfrak{T}}_\a:=t_\alpha + \mathfrak{T}_\alpha\,,\qquad
\check{\mathfrak{S}}_{\a\b}:=s_{\a\b}+ \mathfrak{S}_{\alpha\b}\,,
\end{equation}
the field equations and the energy-momentum and spin laws look even
simpler:
\begin{equation}\label{sum2}
  d\,{\cal F}_\alpha=\check{\mathfrak{T}}_\a\,,\quad  d({\scriptstyle \frac{1}{2}}
  \eta_{\a\b})=\check{\mathfrak{S}}_{\a\b}\,\qquad\text{with}\qquad d\,
  \check{\mathfrak{T}}_\a=0\,,\quad d\, \check{\mathfrak{S}}_{\a\b}=0\,.
\end{equation}
The conservation equations are again implied by the Poincar\'e lemma.

\subsection{Hodge dual of the Freud two-form in Einstein-Cartan
  theory}

As we saw in the definition of the Freud 2-form in a Riemann-Cartan
space in \eqref{Freud2-form'}, the formula looks exactly like its
Riemannian equivalent, only the tilde got lost. Accordingly, the
computation of the Hodge dual in Einstein-Cartan theory exactly parallels
the one in a Rieamannian space. We find again the compact formula
\begin{equation}\label{HodgeF1EC}
  ^\star{\cal F}_\a= \frac{1}{2} \eta_{\a\b\g\d}\,g^{\b\ve}
  {\Gamma}_{\zeta\ve}{}^{\gamma}\eta^{\zeta\d}\,,
\end{equation} 
the tilde is now missing. In turn, the components of \eqref{HodgeF1EC}
read
\begin{equation}\label{semiEC}
 {}^\star{\cal F}^{\m\n}{}_{\a}={
\frac{1}{2}}(\eta_{\a\b\g\d}\, \eta^{\mu\nu\zeta\d})\,g^{\b\ve}
 \, {\Gamma}_{\zeta\ve}{}^{\gamma}={
\frac{1}{2}} \d^{\,\mu\nu\zeta}_{\a\b\g\,}g^{\b\ve}
 \,{\Gamma}_{\zeta\ve}{}^{\gamma}\,.
\end{equation}
If we use again the determinant representation \eqref{determinant} of
the Kroneckers, we find
\begin{equation}\label{FreudDualEC}
 {}^\star{\cal F}^{\m\n}{}_{\a}={
\frac{1}{2}}
\left|
\begin{array}{ccc}
\d^\m_\a   &\d^\n_\a   &\d^\zeta_\a  \\
g^{\m\ve}   &g^{\n\ve}   &g^{\zeta\ve}  \\
 {\Gamma}_{\zeta\ve}{}^{\m}  & 
{\Gamma}_{\zeta\ve}{}^{\n}   & 
{\Gamma}_{\zeta\ve}{}^{\zeta}  
 \end{array}\right|\,.
\end{equation}
This is the generalization of Freud's result to a Riemann-Cartan spacetime.

\subsection{Freud in metric-affine gravity}

Let us finally turn to metric-affine gravity, see
\cite{PRs}. We will denote densities by Gothic letters. The Freud
2-form density and its diamond dual are then defined as
\begin{eqnarray}\label{maFreud} { F}_\a&:=& \frac{1}{2}\,
\mathfrak{g}^{\g\d}\epsilon_{\a\b\g} \wedge\Gamma_\d{}^\b\,,\\ 
\label{maDualFreud}
      \mathfrak{A}_\a&:=&{}^\diamond\! F_\a= \frac{1}{2}\,\mathfrak{g}^{\g\d}\,^\diamond\!\left(\epsilon_{\a\b\g}
\wedge\Gamma_\d{}^\b\right)= \frac{1}{2} \,
\mathfrak{g}^{\g\d}\epsilon_{\a\b\g\ve}\,^\diamond\!\left(\vt^\ve
\wedge\Gamma_\d{}^\b\right)
\nonumber \\
            &=& \frac{1}{2} \,\epsilon_{\a\b\g\ve}\,
            \mathfrak{g}^{\g\d}\,\Gamma_{\zeta\d}{}^\b
           \,^\diamond\!\left(\vt^\ve\wedge\vt^\zeta\right)=
           \frac{1}{2} \,\epsilon_{\a\b\g\ve}\,
           \mathfrak{g}^{\g\d}\,\Gamma_{\zeta\d}{}^\b
\underbrace{\eps^{\ve\zeta}}_{\text{2-form}}\,.
\end{eqnarray}
Let us now determine the components of the 2-form $\mathfrak{A}_\a$. For
the 2-form $\eps^{\ve\zeta}$, we have
$\eps^{\ve\zeta}=\frac 12\,\eps^{\ve\zeta\mu\nu}
\vt_\mu\wedge\vt_\nu$.
Thus,
$\mathfrak{A}_\a=\frac 12\, \mathfrak{A}^{\mu\nu}{}_\a\vt_\m\wedge\vt_\n$ or
\begin{eqnarray}
  \mathfrak{A}^{\mu\nu}{}_\a= \frac{1}{2}\,
  \epsilon_{\a\b\g\ve}\, \eps^{\ve\zeta\mu\nu}
  \mathfrak{g}^{\g\d}\,\Gamma_{\zeta\d}{}^\b= \frac{1}{2}\,
  \delta_{\a\b\g}^{\m\zeta\n}\,
  \mathfrak{g}^{\g\d}\,\Gamma_{\zeta\d}{}^\b\,.
\end{eqnarray}
Our final result,
\begin{eqnarray}\label{semiMAG} \mathfrak{A}^{\mu\nu}{}_\a=
  {
\frac 12}\,
\delta_{\a\b\g}^{\m\n\zeta}\, \mathfrak{g}^{\b\d}\,\Gamma_{\n\d}{}^\g={
\frac{1}{2}}
\left|
\begin{array}{ccc}
\d^\m_\a   &\d^\n_\a   &\d^\zeta_\a  \\
g^{\m\ve}   &g^{\n\ve}   &g^{\zeta\ve}  \\
 {\Gamma}_{\zeta\ve}{}^{\m}  & 
{\Gamma}_{\zeta\ve}{}^{\n}   & 
{\Gamma}_{\zeta\ve}{}^{\zeta}  
 \end{array}\right|\,,
\end{eqnarray}
coincides with the corresponding Riemannian result in
\eqref{semi}. Observe, however, that here we are in the most general
metric-affine space. It seems very likely to us that Freud was aware
of this generalization. Recall that he even called his 2-form
$\mathfrak{A}$ affine tensor density
(``Affintensordichte'').\footnote{The historically genuine Freud
  2-form is then really $\mathfrak{A}_\a$.  But convetionally, it is
  ${\cal F}^\a$ that is named after Freud, see also Mielke
  \cite{Mielke:1992te}, for example.}

If we take the connection in a metric-affine space and decompose it,
we find Schouten \cite{Schouten:1954}, see also \cite[Sec.3.10]{PRs}
and \cite[Eq.(2.132)]{Boehmer},
\begin{eqnarray}\label{MAconnection}
\Gamma_{\alpha\beta} &=& \frac{1}{2} dg_{\alpha\beta} +
   (e_{[\alpha}\rfloor dg_{\beta ]\gamma})\vartheta^{\gamma}
   +  e_{[\alpha}\rfloor C_{\beta ]}
   -\frac{1}{2}(e_{\alpha}\rfloor e_{\beta}\rfloor
   C_{\gamma})\vartheta^{\gamma}
   \hspace{13pt}(\text{Riemann})\nonumber\\
  & &\quad\qquad\qquad\qquad\qquad\quad
  -  e_{[\alpha}\rfloor T_{\beta ]}
  +\frac{1}{2}
  (e_{\alpha}\rfloor e_{\beta}\rfloor
  T_{\gamma})\vartheta^{\gamma}
  \hspace{4.5pt}(\text{Rie-Cartan})\nonumber\\
  &+& \frac{1}{2} Q_{\alpha\beta} +
  (e_{[\alpha}\rfloor Q_{\beta ]\gamma})\vartheta^{\gamma}\, 
  \hspace{110pt} (\text{metric-affine})\,.
\end{eqnarray}
Here $C_\a$ is the object of anholonomity 2-form and $Q_{\a\b}$ the
nonmetricity 1-form. By substituting \eqref{MAconnection} into
\eqref{maFreud} and \eqref{maDualFreud}, we can find the expressions
${\cal F}_\a$ and $\mathfrak{A}_\a$ for the corresponding geometries. Of course,
Freud's result is exactly recovered in the case of Riemannian
geometry.

\section{Conclusions and discussion}

We revisited some well-known results on the superpotential originally
introduced by Freud. By approaching this subject using exterior
calculus, the definition of the superpotential becomes natural and
straightforward, while the usual tensor calculus approach lacks
clarity. In doing so, we were able to extend these results to the
most general case of metric-affine theories of gravity where the
connection may also depend on torsion and nonmetricity.

When rewriting the gravitational field equation using the
superpotential, the equations take the form of a Yang-Mills type
theory. This statement holds for all metric-affine theories and hence
includes general relativity. The only drawback of this formulation is
that it is based on pseudotensors rather than tensors. While the
complete field equations transform correctly under coordinate
transformations, the individual parts do not. A similar issue is
encountered when considering the teleparallel formulation of general
relativity. There the theory is invariant under local Lorentz
transformation, however, many of the individual terms appearing in the
theory are not. It would be interesting to apply our result within the
teleparallel framework.


\end{document}